\documentclass[11pt]{article}
\usepackage{DGfest,epsfig}
\usepackage{times}
\usepackage{psfrag}

\def\be{\begin{equation}}
\def\ee{\end{equation}}
\def\bea{\begin{eqnarray}}
\def\eea{\end{eqnarray}}

\newcommand{\bra}{\langle}
\newcommand{\ket}{\rangle}
\newcommand{\R}{{\cal R}}

\begin{document}
\baselineskip 11.5pt
\title{WHERE IS THE CONFINING STRING IN RANDOM PERCOLATION}

\author{ FERDINANDO GLIOZZI}
\address{ Dipartimento di Fisica Teorica,
        Universit\`a di Torino, and \\
       I.N.F.N., Sezione di Torino, via P.Giuria 1, 
       I-10125 Torino, Italy}  

\maketitle\abstracts{
The percolating phase of whatever random percolation process resembles 
the confining vacuum of a gauge theory in most respects, with a string tension 
having a well-behaved continuum limit, a non trivial glueball spectrum and 
a deconfinement transition at a well determined temperature $T_c$. 
Simple numerical experiments reveal an underlying, strongly fluctuating,  
confining string, with an internal vortex structure 
formed by a core trapping inside a Coulomb-like phase composed by the  
vacuum at the percolation threshold. The width of the core almost 
coincides with $1/T_c$ and it turns out to be separated form the 
confining vacuum by a domain wall of definite thickness.}

\section{Introduction}

Recently, it has been pointed out that three-dimensional percolation can be 
thought as a gauge theory in disguise\cite{Gliozzi:2005ny}. Although 
percolation is a purely geometrical process in which no dynamics is involved,
it captures all the salient features of a confining vacuum.

 Take for 
instance a bond percolation process. The ensemble of configurations
is obtained simply by populating each of the links of a 3D lattice 
independently with occupation probability $p$. The connected graphs 
made by occupied links are called clusters. When $p$ exceeds a threshold 
value $p_c$ there 
is a percolating cluster in the infinite lattice. 

The key ingredient to 
extract from this ensemble the relevant  information on the underlying gauge 
theory is the definition of the percolation counterpart of the Wilson operator
$W_\gamma$, that we associate to whatever loop $\gamma$ of the dual lattice. 
The rule of the game is very simple. We set $W_\gamma=1$ if there is 
no cluster of the 
configuration  topologically linked to $\gamma$, otherwise we put 
$W_\gamma=0$. Therefore, 
the vacuum expectation value $\bra W_\gamma\ket$ coincides with the average 
probability that there is no path in any cluster linked to $\gamma$. 
As in usual gauge theories, evaluating these quantities yields the main 
physical properties of this system. In this way it has been shown that in the 
percolating phase the theory is confining, in analogy to what happens in 
gauge theories with percolating center vortices\cite{'tHooft:1977hy,cp,el}. 
The string tension $\sigma$ and the other physical observables 
have the expected  scaling behaviour dictated by the 
universality class of 3D percolation, therefore such a theory has a well 
defined continuum limit. Moreover it has a non-trivial 
glueball spectrum \cite{Lottini:2005ya} and 
a second-order deconfining transition at finite temperature $T_c$ with 
a ratio $T_c/\sqrt{\sigma}\simeq1.5$ which turns out to be universal, 
i.e. it does not depend  on the kind of lattice utilised nor on the specific 
percolation process considered ( site or bond percolation)
\cite{Gliozzi:2005ny}. 

A particularly 
interesting feature is the occurrence of  universal shape effects in the 
Wilson loops\cite{Gliozzi:2003mb} which are usually ascribed to the 
quantum fluctuations of a string. 
Then the question arises: where and how are encoded the relevant degrees 
of freedom of the underlying confining string? This is the main issue 
discussed in this paper. The answer we find, though surprising, confirms 
once again that percolation is a full-fledged gauge theory: the 
confining string reveals  a non-trivial 
microscopic structure which is believed to be shared by whatever confining 
gauge theory, according to the picture of the confining vacuum as a dual 
superconductor \cite{tm,DiGiacomo:1999fa}.    

The great advantage of percolation is that its simplicity allows to  
explore regions that are still inaccessible 
to the other gauge systems  from a computational point of view.
 
It turns out that the confining string in percolation
is made  of a core in a Coulomb-like vacuum separated 
from the confining vacuum by a domain wall. The width $\R$ of the core 
coincides almost exactly with the inverse of the deconfining temperature, 
while the thickness of the domain wall is about $\R/2$. This confining 
string is not stuck to the minimal surface bounded by $\gamma$, but 
strongly fluctuates, sweeping a large volume which grows with the size 
of the Wilson loop.

\section{Wilson loop as a source of the flux tube}
Each bond configuration of a percolating process in a cubic lattice can be 
mapped into a plaquette 
configuration of the dual lattice by setting empty any plaquette orthogonal to 
an occupied dual link and vice versa. This mapping is one to one, therefore 
the question whether there is a closed path in a cluster that is linked 
with a loop $\gamma$ of the dual lattice is equivalent to the existence 
of a surface $\Sigma$ (at least) of occupied plaquettes having $\gamma$ 
as boundary, i.e. $\partial\Sigma=\gamma$. 
The cluster of this kind of 
surfaces can be considered as the outgoing flux of the ``gauge'' field 
generated by a point-like source describing the world-line $\gamma$. 
\begin{figure}
\begin{center}
\epsfig{figure=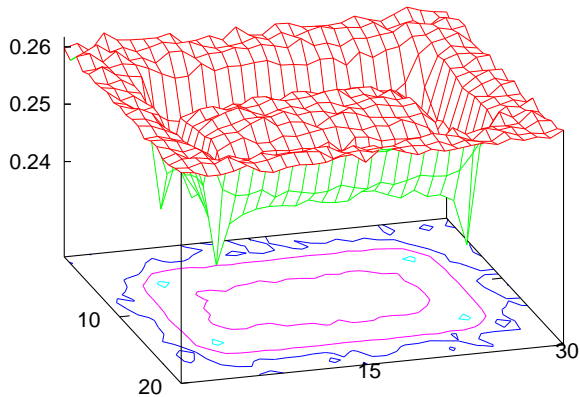,height=2.8in}
\caption{Distribution of the average occupancy probability in the plane of a rectangular $10\times20$ Wilson loop at $p=0.26$.
\label{fig:surface}}
\end{center}
\end{figure}

The crucial observation which allows to extract a lot of information form 
the apparent white noise of the percolation ensemble is that not all the 
configurations are compatible with the existence of such a source, but 
only those with $W_\gamma=1$. In this subset the distribution of occupied 
plaquettes is not  at all flat and may be used to study the spatial 
distribution of the gauge field.
When $p$ is below the percolating threshold the flux is not very constrained 
and is spread out in the whole space. 
On the contrary, in presence of a percolating cluster  the volume at the 
disposal of the gauge flux is much more tight, thus the percolating cluster 
acts as a sort of superconducting medium, squeezing the gauge flux in a 
thin tube.

\begin{figure}
\begin{center}
{\psfrag{p}{$p$}
\psfrag{pc}{$p_c=0.2488126$}
\psfrag{x}{$x$}
\epsfig{figure=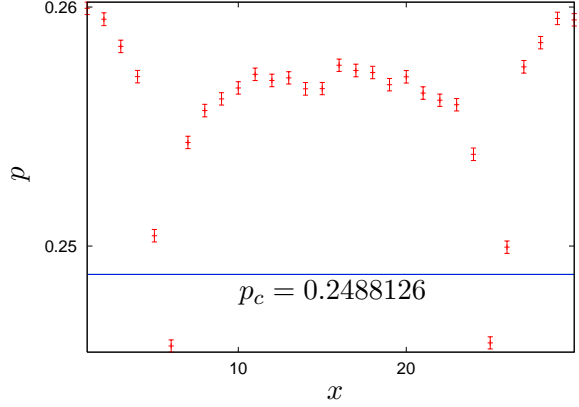,height=2.2in}
}
\caption{Profile of the surface drawn in the previous figure along the 
line at $y=10$. The horizontal straight line denotes the threshold value of 
bond percolation.
\label{fig:profile}}
\end{center}
\end{figure}
It is useful to associate to each link $\ell$ of the lattice a projector $\pi(\ell)$ which is 1 when $\ell$ is occupied an 0 otherwise. We have trivially 
$\bra\pi(\ell)\ket_{E_p}=p$, where $E_p$ is the whole ensemble of the percolation process. Much less trivial is  the average occupancy probability of the links in the subset of configurations with $W_\gamma=1$, that we denote as
\begin{equation}  
\pi_\ell^\gamma\equiv\bra\pi(\ell)\ket_{\{W_\gamma=1\}}~,
\label{pi}
\end{equation}
and we use it as a local probe to explore  the spatial distribution of the 
flux. As an example, the quantity $\pi_\ell^\gamma$ as a function of the 
links $\ell$ 
orthogonal to the plane of a $10\times20$ rectangular Wilson loop in 
the percolating phase  
at $p=0.260$  is drawn in Fig.\ref{fig:surface}. The data are taken using 
a simple burning epidemic type of algorithm \cite{ziff}.  The sinking along the 
perimeter of the rectangle is an effect caused by the  finite clusters: 
since the clusters near $\gamma$ have a finite probability to be linked to it, 
the configurations selected by  $W_\gamma=1$ have a cluster density smaller with respect the whole ensemble $E_p$
\footnote{Also the deepening at the corners of the rectangle is a finite 
cluster effect, because the linking probability of a cluster is 
proportional to its overlap with $\gamma$ and such an overlap is 
larger near a corner.}. Therefore an observer 
near $\gamma$ experiences an occupation probability $\pi^\gamma<p$. 
One has even $\pi^\gamma<p_c$, as  Fig.\ref{fig:profile} shows. 
Hence \emph{the vacuum around the $\gamma$ source is in a 
non-percolating phase}. This important fact will be the starting point of the next section.

Inside the rectangular surface encircled by $\gamma$ the average probability 
$\pi^\gamma_\ell$ reaches a plateau $\tilde p<p$ because of the presence 
of the infinite, percolating cluster. Actually
this very condition is sufficient to assure that 
the vacuum in which the Wilson loop is embedded is confining. 
Indeed it is easy to prove the following \underline{exact} inequality for 
the string tension 
\begin{equation}
\sigma\geq\log\left[\left(\frac{\tilde p}p\right)^{\tilde p}
\left(\frac{1-\tilde p}{1-p}\right)^{1-\tilde p}\right]
\label{string}
\end{equation}      
To prove it, note that the probability  of finding 
$k$ occupied  links among  the $N$ links orthogonal to a rectangle 
$\gamma=r\times t$ is binomial by construction
\footnote{With a similar reasoning one could  ``prove'' quark confinement out 
of percolation of center vortices\cite{el}. There is however a flaw in the argument, because  center vortices, contrary to the  links of random percolation, 
are correlated, therefore they cannot obey a binomial distribution.}:  
:
\begin{equation}
P_N(k)=
\left(
\begin{array}{c}
N \\
k
\end{array}
\right)
p^k(1-p)^{N-k}~.
\end{equation} 
In our case  $N=rt$ and we are assuming that $k\simeq\tilde p \,N$ 
with $\tilde p<p$.  For large $k$ and $N$ we have
\begin{equation}
P_N(\tilde p N)\simeq \frac1{\sqrt{2\pi\tilde p(1-\tilde p)N}}
\left[\left(\frac{\tilde p}p\right)^{\tilde p}
\left(\frac{1-\tilde p}{1-p}\right)^{1-\tilde p}\right]^{-N}.
\label{stirl}
\end{equation}
On the other hand,  not all the configurations with 
$\tilde p\simeq \frac kN$
are compatible with  $W_{r,t}=1$, of course, hence we get the exact inequality 
\begin{equation}
\bra W_{r,t}\ket\leq P_{N}(\tilde{p}\, N)~,~~N=r\,t~,
\end{equation}
which yields, in turn, eq.(\ref{string}).
 
\section{The intrinsic width of the flux tube}  
     
\begin{figure}
\begin{center}
{\psfrag{cv}{confining vacuum}
\psfrag{cp}{Coulomb phase}
\epsfig{figure=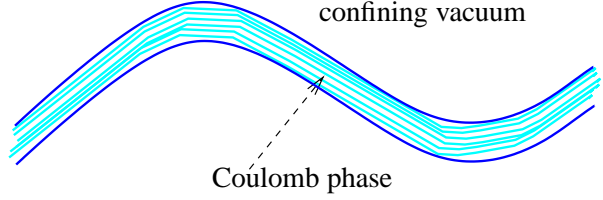,height=1.0in}}
\caption{A  sketch of the microscopic structure of the flux tube. The Coulomb 
phase inside is made of  vacuum at the percolation threshold $p_c$ , 
while the outside vacuum  is at $p>p_c$. 
\label{fig:sigar}}
\end{center}
\end{figure}
What exactly is the confining string made of? 
The distribution of the occupied links  in  close proximity of the source
$\gamma$ of the flux tube is much smaller than its asymptotic value and 
even smaller than the percolation threshold, as Fig. 
\ref{fig:surface} and \ref{fig:profile} clearly show. 
This support the idea that the confining string may have a microscopic 
structure similar  to that of a (dual) Abrikosov or Nielsen-Olesen 
vortex\footnote{For an updated discussion on this argument and a 
complete list of references see \cite{Bolognesi:2005ty}.}, with a core made 
of a Coulomb-like vacuum as contrasted with the surrounding medium, which is in the confining phase. This core  conveys the whole flux in a tube of small but 
non-zero thickness $\R_c$ (see Fig.\ref{fig:sigar}).
Some numerical and analytical tests of this scenario were performed long ago
on the $Z_2$ gauge model \cite{Gliozzi:1996ij,Gliozzi:1996fy}. 
The semi-classical 
descriptions of these extended objects neglect an important 
quantum property of the confining string. In fact, in order to  
account for the universal shape effects  
of  large Wilson loops $\bra W_\gamma\ket$ one is forced  to assume that 
the string world-sheet belongs to a rough phase, hence it should resemble much 
more closely  a strongly fluctuating fluid interface rather than a static, 
smooth structure. As a consequence, we do not expect to 
find a flux tube stuck to the minimal surface encircled by $\gamma$.
Nonetheless, the average occupancy probability in the whole space would 
provide clues to its internal structure.
In this way, enlarging the size of the lattice, one discovers that the 
confining string sweeps a very large volume 
surrounding  the loop $\gamma$. More precisely, we found 
that the  sum 
\begin{equation}
w=\sum_{\ell\in\Lambda}\pi^\gamma_\ell
\label{v}
\end{equation}
over  the $n$  links $\ell$ of the 
lattice $\Lambda$ is less than the naively expected value $p\,n$ and 
their difference $p\,n-w$ approaches a constant value depending only on the 
size of $\gamma$. 
This can be considered as a first hint on the finiteness of the world-volume 
of flux tube. We start by considering a rather crude approximation 
 of the flux tube, by partitioning the $n$ links of the 
lattice into two subsets 
\begin{equation}
n=n_i+n_e~~,
\label{sum}
\end{equation}
where $n_i$ is the number of links lying inside the Coulomb-like core 
of the string and $n_e$ the number of those  lying outside, 
in the confining vacuum.  

The average number of occupied links in the confining phase is 
$p\,n_e$, while the links belonging to the core are occupied with a 
smaller probability $\hat p\leq p_c$ in order to prevent linking. This fact 
is manifest near the fixed boundary $\gamma$, where the string cannot 
fluctuate too much, as shown in Fig.\ref{fig:profile}. Thus we can rewrite
eq.(\ref{v}) as
\begin{equation}
w=\hat{p} \,n_i+p\,n_e~,
\label{ww}
\end{equation}
which tells us that the  volume of the core can be expressed simply as
 \begin{equation}
n_i=\frac{p\,n-w}{p-\hat{p}}~.
\label{volume}
\end{equation}
In order to estimate $n_i$ we need to know the value of $\hat p$. 
A simple scaling argument  suggests $\hat p=p_c$. When $p\to p_c$, 
the flux tube is expected to grow because the infinite,
percolating cluster is crumbling, then the flux can spread out in the whole 
space. Therefore, from one hand we expect in this limit $n_i\sim n$. On 
the other hand in this very limit the confining string disappears, 
hence we should have $w\sim p\,n$. According to eq.(\ref{volume}), these two 
conditions are consistent only if $\hat p=p_c$. 

\begin{table}[t]
\caption{The intrinsic width $\R_c$ of the flux tube generated by a square Wilson 
loop of side $L$. The data are taken at two different values of $p$. The corresponding values of $\sigma$ and of the inverse deconfinement temperature $1/T_c$ are also reported. All the physical quantities are expressed in lattice spacing units. \label{tab:data}}
\vspace{0.4cm}
\begin{center}
\begin{tabular}{|c|c|c|c|c|}
\hline
p&$\sigma$&$1/T_c$& $L$&$\R_c$\\
\hline
&&&10&15.4(3)\\
\cline{4-5}
0.26000& 0.00340(5)&11.5(1)&15&14.8(3)\\
\cline{4-5}
&&&20&14.2(3)\\
\hline
&&&12&12.2(5)\\
\cline{4-5}
0.26502& 0.00649(16)&8.4(1)&16&10.0(6)\\
\cline{4-5}
&&&22&10.6(9)\\
\hline
\end{tabular}
\end{center}
\end{table}

\begin{figure}
\begin{center}
{\psfrag{g}{$\gamma$}
\psfrag{L}{$L_\perp$}
\epsfig{figure=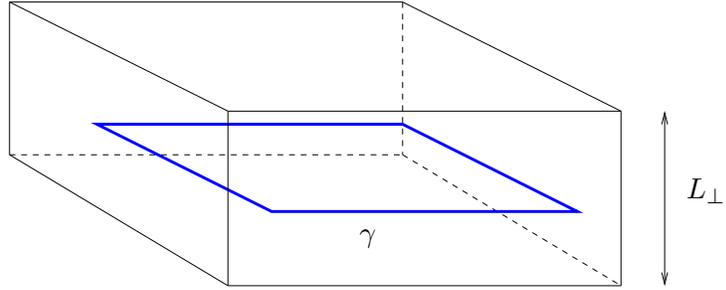,height=1.5in}}
\caption{The slab geometry to probe the internal structure of the flux tube.
The transverse direction is periodic and of size $L_\perp$. $\gamma$ denotes 
a rectangular Wilson loop orthogonal to the compactified dimension.
\label{fig:finte}}
\end{center}
\end{figure}

\begin{figure}
\begin{center}

{\psfrag{Tc}{$1/T_c$}
\psfrag{Rc}{$\R_c$}
\psfrag{L}{$L_\perp$}
\psfrag{R}{$R$}
\epsfig{figure=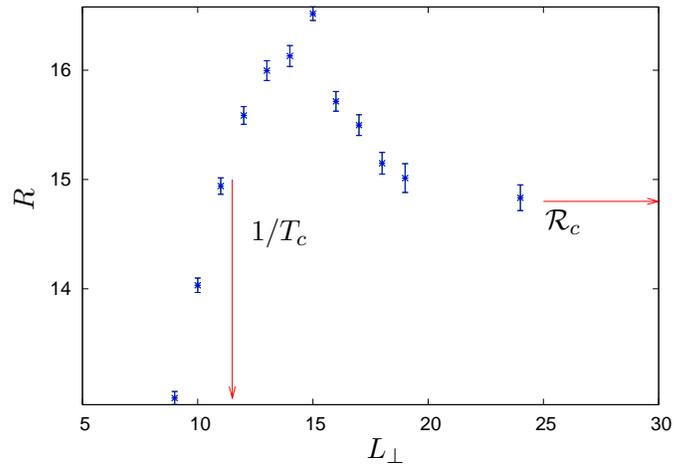,height=2.5in}}
\caption{The intrinsic width of the confining string as a 
function of the transverse dimension $L_\perp$ of the periodic slab 
in which is trapped. The data are taken at $p=0.26$ with a spatial 
square Wilson loop of size $\ell=15$. The horizontal arrow denotes 
the asymptotic value $~R_c(\infty)~$ of the intrinsic width.
\label{fig:compress}}
\end{center}
\end{figure}

Assuming the world-volume of the 
flux tube to be a slab of size $A\,\R_c$, with $A$ the area of the minimal 
surface encircled by $\gamma$, we are led to the following effective definition
of the intrinsic thickness of the flux tube 
\begin{equation}
\R_c=\frac{n_i}{A}= \frac{p\,n-w}{A(p- p_c)}~.
\label{rc}
\end{equation}
Some numerical estimates 
of $\R_c$  are reported  in Tab.\ref{tab:data}. They seem to approach 
asymptotically an approximate scaling behaviour, as expected for a 
physical quantity. Though it gives a length scale of the order of the 
transverse size of the flux tube, it should not be taken too literally. First,
the splitting of the links in two subsets of those inside or outside the 
Coulombic core is an over-simplification: there should be a domain wall between the two vacua with an average occupancy probability that interpolates between 
$p$ and $p_c$. This would produce an enlargement of the flux tube which is not taken into account in eq.(\ref{rc}). 
Similarly, the number of interior links does not coincide exactly 
with the volume of the core, the basis of the 
slab occupied by the world-volume of the 
flux tube is actually larger than $A$, and so on. Nevertheless we can still 
extract further information on the microscopic structure of the confining 
string combining these results with another kind of numerical experiments.

Long time ago it was observed that in three-dimensional  $SU(2)$ 
\cite{mt} and $Z_2$ gauge models\cite{Caselle:1993cb,Caselle:1993mt}
the intrinsic width of the flux tube almost coincides with the inverse of the deconfinement temperature (see also \cite{mey}). This result was obtained by 
studying the response of the vacuum expectation value of a large, planar 
Wilson loop when one varies the size $L_\perp$ of the dimension orthogonal to 
the  loop in a periodic lattice, like in Fig.\ref{fig:finte}. 
\begin{figure}
\begin{center}
{\psfrag{W}{$\bra W_{\ell,\ell}\ket$}
\psfrag{L}{$L_\perp$}
\epsfig{figure=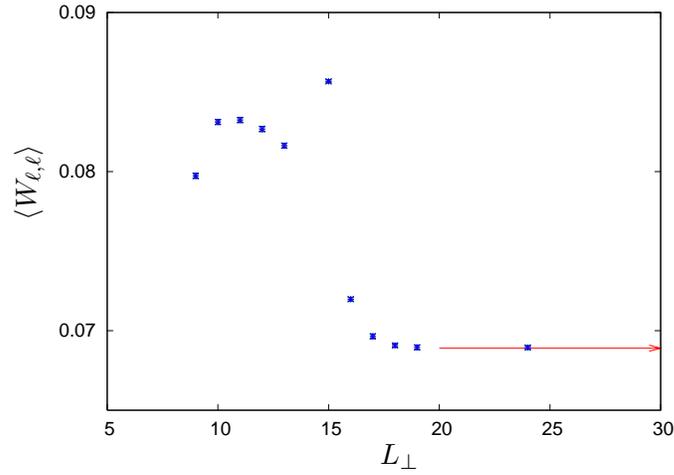,height=2.5in}}
\caption{Vacuum expectation value of a square Wilson loop of size $\ell=15$
at $p=0.26$ as a function of the  transverse periodic 
dimension $L_\perp$ of the lattice. The arrow indicates its 
asymptotic value.
\label{fig:wilc}}
\end{center}
\end{figure}

We repeated 
such a numerical experiment in percolation, using the same setting and 
measuring the intrinsic radius $\R_c$ as defined in eq.(\ref{rc}). 
The surprising results are reported in Fig.\ref{fig:compress} and 
\ref{fig:wilc}. Let us follow  the behaviour of $R_c(L_\perp)$ starting from 
$L_\perp$ large enough, where $R_c$ coincides with its asymptotic value
$\R_c\equiv R_c(\infty)$. As $L_\perp$
decreases, we observe first an apparent growth of $R_c(L_\perp)$ starting at a 
value of $L_\perp$  slightly larger than the asymptotic value 
$\R_c$, the reason being perhaps the underestimated contribution 
of the domain wall. The apparent increasing of  $R_c(L_\perp)$ should be 
ascribed to the fact that the two domain walls separating the core from the 
confining vacuum are pushed one against the other squeezing a fraction 
of the core. When $L_\perp$ is of the order of $1/T_c$, the external 
confining vacuum disappears and the 
 domain walls melt in the  Coulomb-like core.  At this point the flux 
tube fills the entire interval $L_\perp$ and  loses its own width, which 
now coincides, for $L_\perp\leq1/T_c$, with the transverse size of 
the lattice. This fact is corroborated by value of the slope in this region, 
which is about 1.

In conclusion, the response of the system to the variation of the size of the 
transverse dimension seems to confirm the scenario of a confining string 
 made  of a core in a Coulomb-like vacuum separated 
from the confining medium by a domain wall. The width $\R$ of the core 
almost coincides with the inverse of the deconfining temperature, like 
in other gauge systems, while the thickness of the domain wall seems to be 
about $\R/2$.

\end{document}